\documentclass[prb,twocolumn,reprint,superscriptaddress,showpacs]{revtex4-1}

\usepackage[version=3]{mhchem} 
\usepackage{amssymb} 
\usepackage{textcomp} 
\usepackage{graphicx}
\usepackage[utf8]{inputenc}
\usepackage{xcolor}

\begin{document}

\author{C. M. Polley}
\email{craig.polley@maxlab.lu.se}
\affiliation{MAX IV Laboratory, Lund University, 221 00 Lund, Sweden}

\author{V. Jovic}
\affiliation{School of Chemical Sciences and MacDiarmid Institute for Advanced Materials and Nanotechnology, University of Auckland,
Auckland 1142, New Zealand}

\author{T.-Y. Su}
\affiliation{Department of Physics, Boston University, 590 Commonwealth Avenue, Boston, Massachusetts 02215, USA}

\author{M. Saghir}
\affiliation{Physics Department, University of Warwick, Coventry, CV4 7AL, United Kingdom}

\author{D. Newby, Jr.}
\affiliation{Department of Physics, Boston University, 590 Commonwealth Avenue, Boston, Massachusetts 02215, USA}

\author{B. J. Kowalski}
\affiliation{Institute of Physics, Polish Academy of Sciences, 02-668 Warszawa, Poland}

\author{R. Jakiela}
\affiliation{Institute of Physics, Polish Academy of Sciences, 02-668 Warszawa, Poland}

\author{A. Barcz}
\affiliation{Institute of Physics, Polish Academy of Sciences, 02-668 Warszawa, Poland}

\author{M. Guziewicz}
\affiliation{Institute of Electron Technology, Al. Lotnikow 32/46, 02-668 Warszawa, Poland}

\author{T. Balasubramanian}
\affiliation{MAX IV Laboratory, Lund University, 221 00 Lund, Sweden}

\author{G. Balakrishnan}
\affiliation{Physics Department, University of Warwick, Coventry, CV4 7AL, United Kingdom}

\author{J. Laverock}
\affiliation{Department of Physics, Boston University, 590 Commonwealth Avenue, Boston, Massachusetts 02215, USA}
\affiliation{H. H. Wills Physics Laboratory, University of Bristol, Tyndall Avenue, Bristol, BS8 1TL, UK}

\author{K. E. Smith}
\affiliation{Department of Physics, Boston University, 590 Commonwealth Avenue, Boston, Massachusetts 02215, USA}
\affiliation{School of Chemical Sciences and MacDiarmid Institute for Advanced Materials and Nanotechnology, University of Auckland,
Auckland 1142, New Zealand}

\title{Observation of surface states on heavily indium doped SnTe(111), a superconducting topological crystalline insulator}

\begin{abstract}

The topological crystalline insulator tin telluride is known to host superconductivity when doped with indium (Sn$_{1-x}$In$_{x}$Te), and for low indium content ($x=0.04$) it is known that the topological surface states are preserved. Here we present the growth, characterization and angle resolved photoemission spectroscopy analysis of samples with much heavier In doping (up to $x\approx0.4$), a regime where the superconducting temperature is increased nearly fourfold. We demonstrate that despite strong p-type doping, Dirac-like surface states persist.

\end{abstract}


\maketitle

\section{Introduction}

Three dimensional topological insulators constitute a popular and active topic in contemporary condensed matter physics, primarily stemming from the exotic properties of the surface states in these systems. This set of properties can be enriched even further when combined with superconductivity \cite{Fu2008,Beenakker2013,Fu2015}. Notably, such a system represents a quasiparticle approach to realizing a Majorana fermion. While such a realization is of considerable importance to the particle physics community, the interest also extends further into quantum computation, where Majorana bound states potentially represent a qubit platform.

Accordingly, there is an active branch of research into candidate materials. One approach is to deposit a superconductor on a regular TI, relying on the proximity effect \cite{Fu2008,Xu2014}. An alternative is to find bulk-superconducting materials which natively possesses topological surface states - examples are Cu$_x$Bi$_2$Se$_3$ (T$_C=3.8$K for $x\approx0.14$) \cite{Hor2010,Sasaki2011,Das2011} and Sn$_{1-x}$In$_{x}$Te \cite{Erickson2009,Sasaki2012,He2013,Zhong2013,Balakrishnan2013, Lees2014}. For the latter, an indium content of $x=0.04$ yields a critical temperature of 1.2~K and topological surface states appear to persist \cite{Sato2013}. However it has recently been realised that the critical temperature can be increased to 4.5~K with a higher In content of $x=0.45$ \cite{Zhong2013}. An immediate question is whether this increase comes at the cost of the topological surface states. Specifically, while the lower limit of cubic SnTe is a confirmed topological crystalline insulator, the same cannot be said of the lower symmetry upper limit, tetragonal InTe. 

Angle-resolved photoemission spectroscopy (ARPES) can play an important role here, but the strong p-type doping in SnTe due to natural Sn vacancies \cite{Wang2014_1} makes this challenging. The doping becomes even more severe with the introduction of indium, which acts as a p-type dopant in SnTe \cite{Erickson2009}. To date a Dirac point has never been observed in photoemission measurements of pure SnTe, but can nonetheless be inferred from measurements of heavily lead doped SnTe \cite{Xu2012,Yan2014}. Here we report the \textit{in-situ} growth and photoemission study of (111) oriented Sn$_{1-x}$In$_{x}$Te films with very high indium contents. Despite severe p-type doping, we clearly observe that Dirac-like surface states persist at these high indium contents.

\section{Methods}

High quality (111) oriented surfaces of (Sn,In)Te were prepared in ultrahigh vacuum (UHV) by a hot-wall epitaxy method. After \textit{ex-situ} cleaving shortly before entry into UHV, BaF$_2$ substrates were prepared by being heating to $>$400~\textcelsius{} for several hours. For deposition the substrate temperature was reduced to $\approx$330~\textcelsius{}. Thick films of (Sn,In)Te ($\approx$1.5$\mu$m) were deposited by single source open hot-wall epitaxy \cite{HWE_Review}, using \textit{ex-situ} re-crystallized Sn$_{0.6}$In$_{0.4}$Te in powdered form as source material. All subsequent photoemission measurements were performed without removing the samples from UHV ($<2\times$10$^{-10}$~Torr). LEED measurements (Figure {\ref{fig:corelevels}a) confirm that the (Sn,In)Te films grow with a (111) orientation. 

All UV photoemission measurements were performed at the I4 beam line at the MAX-IV synchrotron facility \cite{Jensen1997}, using p-polarized photons. The total energy resolution (beamline and spectrometer) was configured for 25~meV, except for the higher energy scans in Figures \ref{fig:bandstructure1} and \ref{fig:fsm} ($\approx$85~meV). 
Fermi level positions within the films were determined by reference to a clean tantalum foil on the manipulator. This was found to be accurate to within only $\approx$30~meV in most cases, presumably owing to charging effects from the BaF$_2$ substrate. For high resolution valence band spectra, the Fermi level position was revised by careful examination of the intensity distribution in the spectra.

Owing to its low melting temperature and high vapor pressure relative to Sn, In has a tendency to segregate to the surface during film growth (which we discuss in more detail below). In order to understand the depth-dependent In concentration, complementary hard x-ray photoemission (HAXPES) measurements were also performed on separate thin film samples, as well as on the precursor Sn$_{0.6}$In$_{0.4}$Te single crystals. For these measurements, photon energies of 2.14~keV and 4~keV were used, with energy resolutions of $\approx180$~meV and $\approx220$~meV and probing depths of $\approx3.5$~nm and $\approx6.5$~nm, respectively (compared with $<1$~nm at UV energies). The thin film samples used for the HAXPES experiment were grown and characterized in UHV at Beamline X1B of the NSLS using similar parameters to the UV ARPES samples, before being transferred (with exposure to atmosphere) to the HAXPES instrument. The angle-integrated valence band of both bulk single crystal and as-grown thin film were very similar to one another (Figure {\ref{fig:corelevels}e), and in good agreement with previous measurements of SnTe \cite{Kemeny1976}, confirming that electronically the grown films are In-doped SnTe.

\section{Stoichiometry characterization}

\begin{figure}
	\includegraphics[width=8cm]{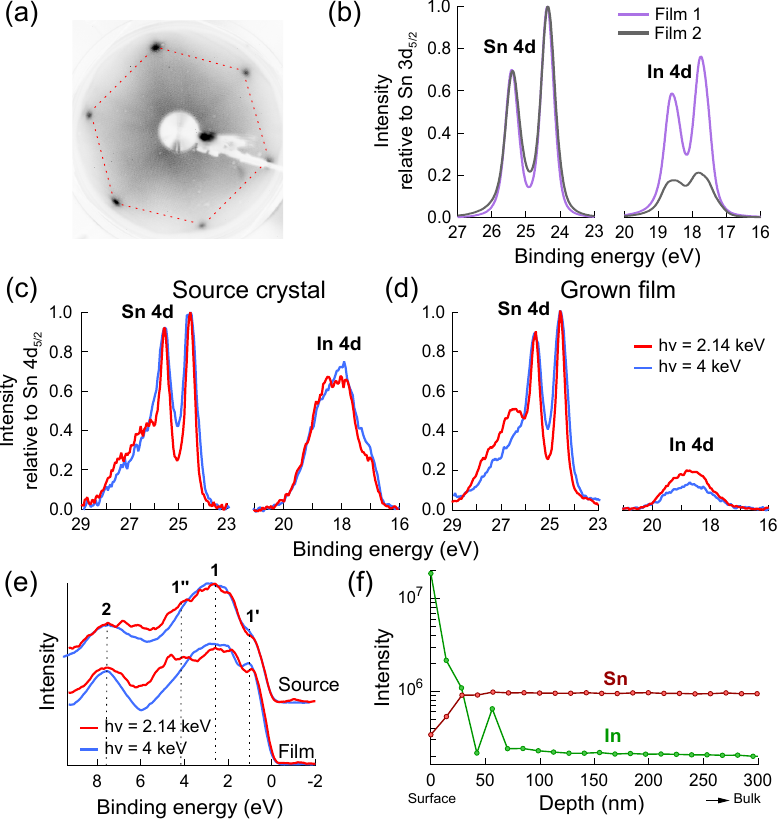}
	\caption{
		\footnotesize{(a) Low energy (55~eV) electron diffraction pattern of a typical film. (b) Normal emission UPS spectra of the 4d core levels of Sn and In for film 1 (h$\nu$=70~eV) and film 2 (h$\nu$=94~eV); both films exhibit a large indium signal. (c-e) The same core levels excited with high energy photons (h$\nu$=2-4~keV) for both the Sn$_{0.6}$In$_{0.4}$Te source material and a third thin film preparation. The reduction of the In 4d peak with increasing photon energy in the thin film is indicative of indium segregation during film growth. The valence band spectra are labelled according to the scheme of Kemeny \cite{Kemeny1976}, and demonstrate that the films grow as In-doped SnTe. (e) SIMS profile of film 1, confirming that indium segregates to the surface.
		}
	}
	\label{fig:corelevels}	
\end{figure}

In this article we will mainly discuss two film samples, produced with slightly different source and sample temperatures and thus with different stoichiometry. Normal emission ultraviolet core level spectroscopy (UPS) (Figure {\ref{fig:corelevels}b) at photon energies of 70~eV and 94~eV suggests an indium content which is significantly higher than the $x=0.045$ bulk crystals studied previously \cite{Sato2013}. A simplistic quantification based on the ratio of peak areas and photoionization cross sections\cite{CrossSectionComment1} suggests surface indium contents of $x=0.58$ and $x=0.33$ for the two films. Similar analysis of a uniform $x=0.40$ bulk crystal at a photon energy of 80~eV yields an indium fraction of $x=0.56$ \cite{Supplementary}, suggesting that in this low energy regime the theoretically calculated cross section ratio should be slightly rescaled. This lowers the estimate for our films to more realistic values of $x=0.41$ and $x=0.23$. Scanning the $\approx$100~$\mu$m$\times$25~$\mu$m light spot across the sample showed no obvious variation in the core level ratios.

Supporting measurements of the film stoichiometry are not trivial owing mainly to indium segregation at the surface, i.e.\ the stoichiometry is strongly depth dependent. This is confirmed by both HAXPES measurements of a separate (but similarly prepared) sample, as well as secondary ion mass spectroscopy (SIMS) of the $x=0.41$ film. Figure \ref{fig:corelevels}c and \ref{fig:corelevels}d compares core level spectra of the grown film and the bulk precursor crystal, measured with photon energies of 2.14~keV ($\lambda_{4d}\approx$3.5~nm) and 4~keV ($\lambda_{4d}\approx$6.5~nm). Here $\lambda_{4d}$ is the inelastic mean free path of the In/Sn $4d$ photoelectrons, a measure of the depth sensitivity. For the $x=0.40$ bulk crystal the core level peak areas are not appreciably different for the two photon energies, consistent with a spatially uniform indium content. The same quantification technique as employed for the UPS spectra yields indium fractions of $x=0.49$ at 2.14~keV and $x=0.48$ at 4~keV. In contrast, for the film the indium core level signal is notably smaller, and appears significantly reduced at higher photon energies, mapping to indium fractions of $x=0.18$ at 2~keV and $x=0.12$ at 4~keV\cite{CrossSectionComment2}. That the indium signal is reduced as $\lambda_{4d}$ is increased suggests a decaying indium concentration, and this is confirmed by SIMS measurements (Figure {\ref{fig:corelevels}e).

For the purposes of interpreting our angle resolved spectra it is only the near-surface stoichiometry that is of principle interest, since the photoelectrons in this case possess very short inelastic mean free paths ($<$1~nm). Techniques such as energy dispersive x-ray spectroscopy (EDX) with probing depths of several $\mu$m are therefore not representative. For example, EDX performed on the films grown in this study yielded indium concentrations as low as x$\approx$0.04, corresponding to the depth-averaged In content and clearly not in agreement with the surface sensitive \textit{in-situ} UPS measurements. Throughout the manuscript we will continue to label the two Sn$_{1-x}$In$_{x}$Te films by UPS estimated indium contents of $x=0.41$ and $x=0.23$, but it should be understood that these values represent only a best estimate. This method of stoichiometry determination can be influenced for example by resonance effects, detector non-linearity or the accuracy of photoionization cross sections.

It is interesting to consider why such a high indium content has apparently been obtained in a single-phase crystal, when previous work has indicated the emergence of a tetragonal InTe phase at high indium contents. A likely answer to this lies again with the strong In segregation, in the sense that we have produced not a large, uniform single crystal with a high indium content but only an In enriched surface layer. It seems reasonable to expect that the phase diagram could be quite different for very thin films.

\section{Electronic structure measurements}
\begin{figure}	
	\includegraphics[width=8cm]{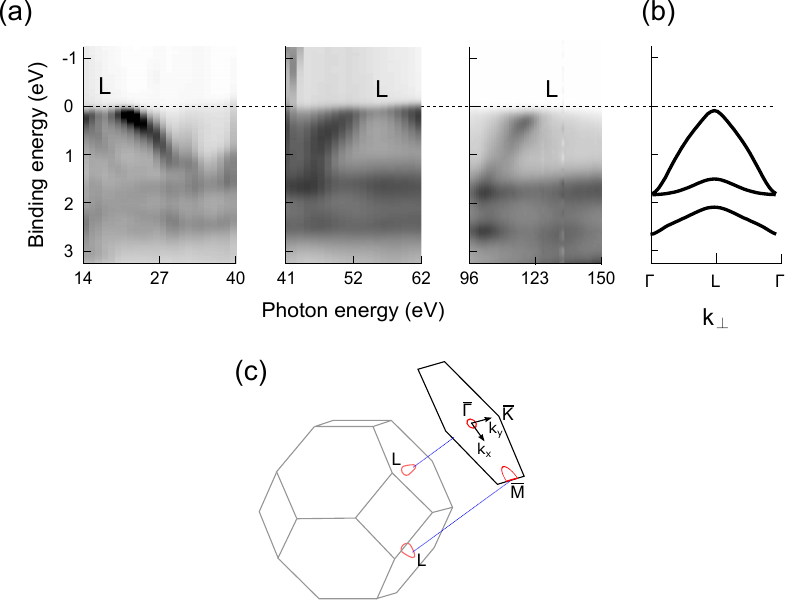}
	\caption{
		\footnotesize{(a) Room temperature, normal emission photoemission spectra for the $x=0.41$ sample as a function of photon energy. Such a measurement maps the k$_Z$ dispersion along the bulk L-$\Gamma$-L high symmetry direction. The experimental dispersions agree well with tight binding calculations for bulk SnTe (b, adapted with permission from Littlewood et al \cite{Littlewood2010}). (c) A schematic depiction of the relationship between the (111) surface and the bulk Brillouin zone illustrates the projection of Fermi ellipsoids at the bulk L-points onto  $\bar{\Gamma}$ and $\bar{M}$ in the surface Brillouin zone.}
	}
	\label{fig:bandstructure1}
\end{figure}

In Figure \ref{fig:bandstructure1} we show normal emission photoemission intensity maps as a function of photon energy for the $x=0.41$ film. The band structure is in good agreement with the $\Gamma$-L-$\Gamma$ direction of bulk SnTe \cite{Littlewood2010}, providing further confidence that we are studying well defined, (111) oriented (Sn,In)Te as opposed to In$_{x}$Te$_{y}$ clusters. At normal emission the bulk L points are probed at photon energies of approximately 17~eV, 55~eV and 113~eV, and around these energies the bulk valence band crosses the Fermi level at $\bar{\Gamma}$. Upon closer inspection (Figure \ref{fig:hvdependence}) it becomes apparent that two distinct pairs of bands are present. The innermost pair broadens and disperses to a higher binding energy as the photon energy is increased, consistent with the bulk valence band. In contrast, the outermost pair does not disperse, a characteristic property of surface states. As is also the case for the surface states of (Pb,Sn)Se(111), the surface state intensity is resonantly enhanced at photon energies which probe the bulk L point \cite{Polley2014}.

\begin{figure}	
	\includegraphics[width=8cm]{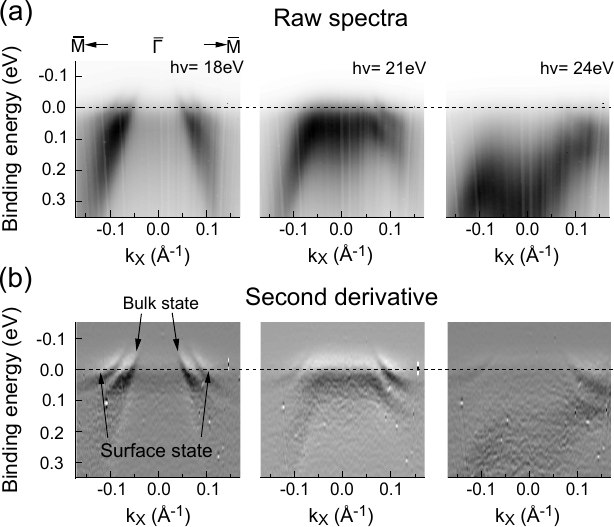}
	\caption{
		\footnotesize{Photon energy dependent ARPES of the hole pockets at $\bar{\Gamma}$ for the $x=0.41$ sample (T$\approx$230~K). Raw spectra (a) and corresponding second derivative images after box smoothing along the energy direction (b). The outermost bands do not disperse with increasing photon energy, identifying them as a surface states. The innermost pair broadens and moves to higher binding energy, consistent with the bulk valence band at bulk L.}
	}
	\label{fig:hvdependence}
\end{figure}

\begin{figure}	
	\includegraphics[width=8cm]{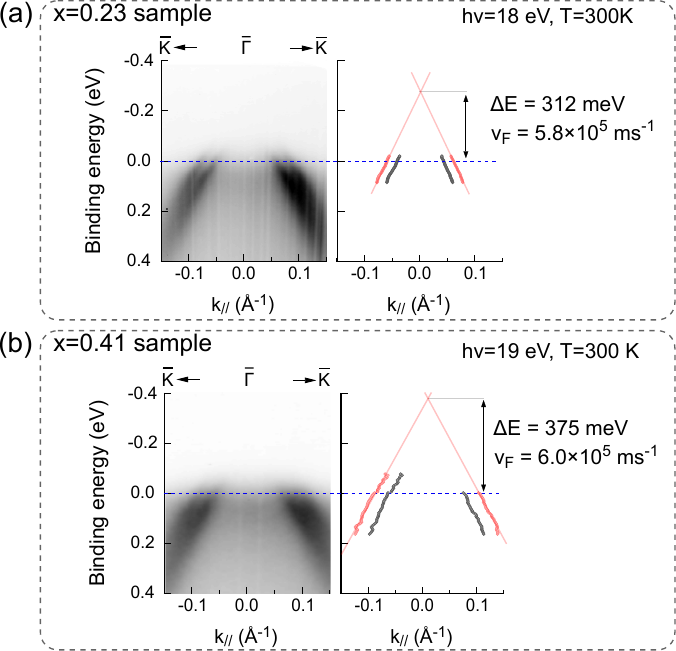}
	\caption{
		\footnotesize{Quantitative dispersion mapping of the $\bar{\Gamma}$ hole pockets for two different indium content samples (a,b). By fitting individual momentum distribution curves, the dispersion of both the bulk and surface bands can be tracked. A linear extrapolation of the surface state bands gives a measure of the doping level and Fermi velocity for the two different samples.}
	}
	\label{fig:doping}
\end{figure}
The dispersion of the surface state resembles the linear dispersion typical of topological surface states close to a Dirac point, and indeed one might anticipate topological states on the basis of previous studies of SnTe \cite{Tanaka2012} and Sn$_{0.955}$In$_{0.045}$Te \cite{Sato2013}. In Figure \ref{fig:doping} we quantify the dispersion, tracking peak positions by fitting individual momentum distribution curves taken along $\bar{K}$-$\bar{\Gamma}$-$\bar{K}$ with four Lorentzian peaks. Linear extrapolation of the surface state bands then provides an estimate of the binding energy of the (assumed) Dirac points: 375~meV above the Fermi level for the $x=0.41$ film and 312~meV for the $x=0.23$ film. Previous studies found values of 50~meV for both SnTe \cite{Tanaka2012} and Sn$_{0.955}$In$_{0.045}$Te \cite{Sato2013}, at least partially due to surface band-bending. The larger value we observe compared to these studies may reflect different surface band-bending, but is also consistent with a significantly higher indium content.

\begin{figure}	
	\includegraphics[width=8cm]{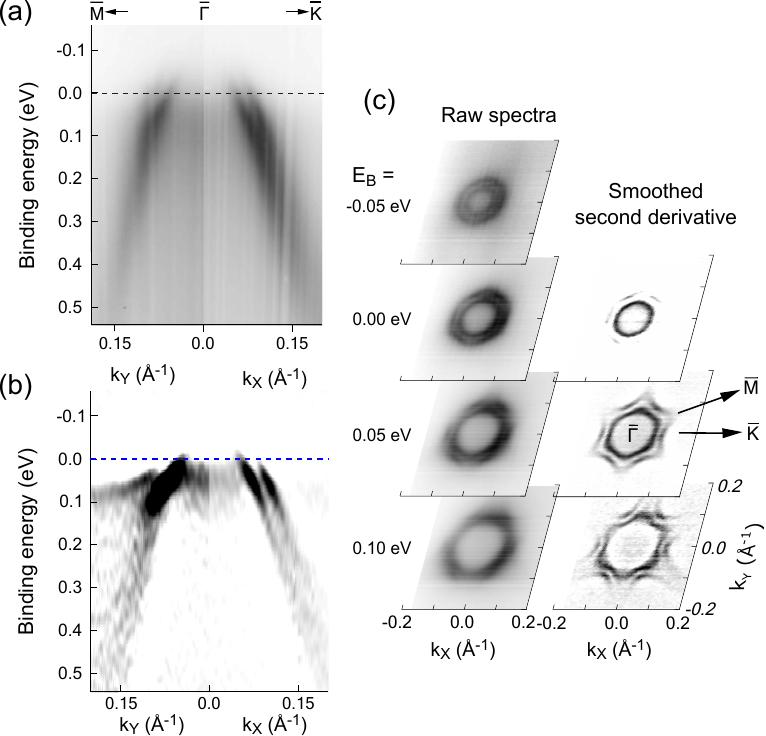}
	\caption{
		\footnotesize{Room temperature three dimensional ARPES mapping of the $\bar{\Gamma}$ hole pockets in the $x=0.23$ film at h$\nu$=18~eV. Above the Fermi level the contours resemble an isotropic 2 dimensional hole gas, but at higher binding energies the surface state becomes severely warped towards $\bar{M}$. This becomes particularly clear in second derivative images (b,c)}
	}
	\label{fig:fsm}
\end{figure}

Constant energy surface mapping of the $\bar{\Gamma}$ surface state (Figure \ref{fig:fsm}) reveals both intensity modulations and strong hexagonal warping, especially apparent in second-derivative images. In Figure \ref{fig:fsm} we present such mapping for the $x=0.23$ film. The $x=0.41$ data is qualitatively similar, but less of the surface state bandstructure is visible due to the heavier doping. The low energy surface electronic structure around $\bar{\Gamma}$ in spin-orbit coupled systems can often be well described by a simple 2D nearly-free-electron model, giving rise to circular Fermi contours. This is at least true of Au(111) and Bi$_{2}$Se$_{3}$ close to the Dirac point, but has been shown to be insufficient for stronger spin-orbit coupled systems such as Bi$_{2}$Te$_{3}$ and the giant Rashba system BiAg$_{2}$ \cite{Frantzeskakis2011,Vajna2012}. In such cases the Hamiltonian must be expanded to include higher order terms in momentum, with the crystal symmetry dictating allowable terms. For $C_{3V}$ symmetric systems such as Sn$_{1-x}$In$_{x}$Te(111) this leads to hexagonal warping and a variety of interesting consequences such as surface density wave instabilities, lifetime anisotropy and an out-of-plane spin texture \cite{Fu2009,Nomura2014,SanchezBarriga2014}. We note that this is the first clear observation of hexagonal warping effects in a topological crystalline insulator system. Previous studies of (111) oriented SnTe and (Sn,In)Te \cite{Tanaka2013, Sato2013} did not sufficiently resolve the Fermi surface, whilst in (111) oriented Pb$_{1-x}$Sn$_{x}$Se \cite{Polley2014} the warping effect is expected to be much less severe owing to the weaker spin-orbit coupling of selenium compared to tellurium (analogous to the comparison of Bi$_{2}$Se$_{3}$ and Bi$_{2}$Te$_{3}$). The combination of warping effects with a superconducting transition makes Sn$_{1-x}$In$_{x}$Te(111) an interesting system for continued studies.

\begin{figure}	
	\includegraphics[width=8cm]{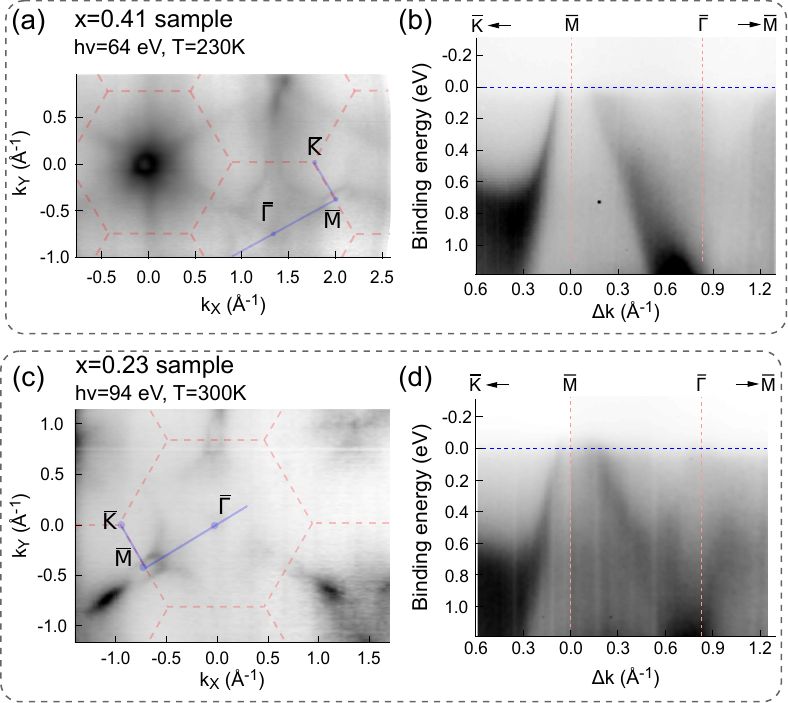}
	\caption{
		\footnotesize{Fermi surface maps covering several surface Brillouin zones of the $x=0.41$ (a) and $x=0.23$ (c) samples. The intensity of different features within the maps is highly non-uniform, but cuts along $\bar{M}-\bar{\Gamma}-\bar{M}-\bar{K}$ in both samples (b,c) indicate anisotropic hole pockets located at $\bar{M}$.}
	}
	\label{fig:sbz}
\end{figure}

Based on consideration of mirror symmetry and where the bulk $L$-points project to, Dirac points are also predicted to exist at $\bar{M}$ (Fig. \ref{fig:bandstructure1}c). However the configuration of these Dirac points depends on the surface termination. In contrast to the well studied (100) natural cleavage plane, the (111) face of SnTe is strongly polar. Different surface terminations and reconstructions are therefore possible, with significant consequences for the surface electronic structure \cite{Eremeev2014,Wang2014_2}. Thermodynamic calculations for SnTe indicate that a $(1 \times 1)$-Te termination is the most stable, in which case the $\bar{M}$ Dirac point is expected to lie close to the bulk conduction band edge \cite{Wang2014_2}. Combined with the strong p-type doping, it is therefore not obvious that Dirac cones should be experimentally observable at $\bar{M}$. This is further complicated by the strong intensity resonances in this system - even if present, surface states can only be resolved at particular photon energies. In Figure \ref{fig:sbz}a,c we show wide-area Fermi surface maps for both the $x=0.41$ sample (h$\nu$=64~eV) and the $x=0.23$ sample (h$\nu$=94~eV). In both cases there are weakly resolved anisotropic hole pockets at $\bar{M}$. In Figure \ref{fig:sbz}b,d we extract energy-momentum spectra from regions where these pockets are most visible. The fact that the linear dispersions appear to be unchanged between the 64~eV and 94~eV scans is suggestive that these are indeed the lower portions of $\bar{M}$ surface state Dirac cones.

\section{Conclusions}
By growing (Sn,In)Te films \textit{in-situ}, we have bypassed the need for cleavable bulk crystals and been able to study the high-indium regime of this material. Importantly, we can confirm that surface states similar in character to the topological states on SnTe continue to exist at very high indium contents. While spin-resolved measurements are always a necessary requirement to confirm the nature of these surface states, our observation is nonetheless highly encouraging for the continuing efforts towards superconducting topological insulators. In particular, we note that the increase in critical temperature for Sn$_{1-x}$In$_{x}$Te when the indium content reaches $x=0.45$ brings it within a highly accessible liquid helium experimental regime. However as our measurements confirm, it remains an important and unsolved material challenge to realize n-type doping of (Sn,In)Te such that transport properties are not dictated by trivial bulk bands. To this end, the possibility of high quality thin-film growth demonstrated here may open up new research directions.


\begin{acknowledgments}
This work was made possible through support from the Knut and Alice Wallenberg Foundation and the Swedish Research Council. Work at the University of Warwick was supported by the EPSRC, UK (Grant EP/L014963/1). The Boston University program is supported in part by the US Department of Energy under Grant No.\ DE-FG02-98ER45680. The National Synchrotron Light Source, Brookhaven, is supported by the US Department of Energy under Contract No.\ DE-AC02-98CH10886. We thank Jacek Osiecki for creating data analysis software used in this study, and Barry Karlin and Joseph Woicik for technical assistance

\end{acknowledgments}


\begin{thebibliography}{99}
\bibitem{CrossSectionComment1} Area quantification was performed by fitting with a spin-orbit split Voigt doublet and Shirley background. The cross sections used were In4d$_{70eV}$=16.1~Mb, In4d$_{94eV}$=2.17~Mb Sn4d$_{70eV}$=23.9~Mb and Sn4d$_{94eV}$=3.26~Mb, obtained from Elettra's``WebCrossSection'' service
\end{thebibliography}
\end{document}


\author{C. M. Polley}
\email{craig.polley@maxlab.lu.se}
\affiliation{MAX IV Laboratory, Lund University, 221 00 Lund, Sweden}

\author{V. Jovic}
\affiliation{School of Chemical Sciences and MacDiarmid Institute for Advanced Materials and Nanotechnology, University of Auckland,
Auckland 1142, New Zealand}

\author{T-Y. Su}
\affiliation{Department of Physics, Boston University, 590 Commonwealth Avenue, Boston, Massachusetts 02215, USA}

\author{M. Saghir}
\affiliation{Physics Department, University of Warwick, Coventry, CV4 7AL, United Kingdom}

\author{D. Newby, Jr.}
\affiliation{Department of Physics, Boston University, 590 Commonwealth Avenue, Boston, Massachusetts 02215, USA}

\author{B. J. Kowalski}
\affiliation{Institute of Physics, Polish Academy of Sciences, 02-668 Warszawa, Poland}

\author{R. Jakiela}
\affiliation{Institute of Physics, Polish Academy of Sciences, 02-668 Warszawa, Poland}

\author{A. Barcz}
\affiliation{Institute of Physics, Polish Academy of Sciences, 02-668 Warszawa, Poland}

\author{M. Guziewicz}
\affiliation{Institute of Electron Technology, Al. Lotnikow 32/46, 02-668 Warszawa, Poland}

\author{T. Balasubramanian}
\affiliation{MAX IV Laboratory, Lund University, 221 00 Lund, Sweden}

\author{G. Balakrishnan}
\affiliation{Physics Department, University of Warwick, Coventry, CV4 7AL, United Kingdom}

\author{J. Laverock}
\affiliation{Department of Physics, Boston University, 590 Commonwealth Avenue, Boston, Massachusetts 02215, USA}
\affiliation{H. H. Wills Physics Laboratory, University of Bristol, Tyndall Avenue, Bristol, BS8 1TL, UK}

\author{K. E. Smith}
\affiliation{Department of Physics, Boston University, 590 Commonwealth Avenue, Boston, Massachusetts 02215, USA}
\affiliation{School of Chemical Sciences and MacDiarmid Institute for Advanced Materials and Nanotechnology, University of Auckland,
Auckland 1142, New Zealand}

\title{Supplementary information for `Observation of surface states on heavily indium doped SnTe(111), a superconducting topological crystalline insulator'}

\maketitle
\begin{figure}
	\includegraphics[width=6.3cm]{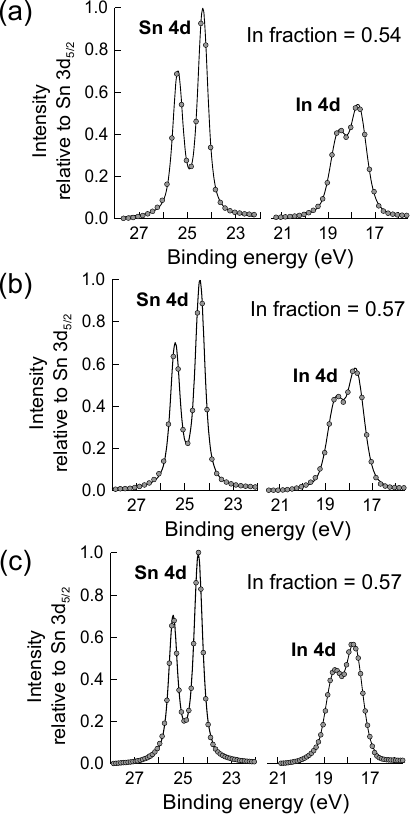}
	\caption{
		\footnotesize{Normal emission UPS spectra (h$\nu$=80~eV) of the 4d core levels of Sn and In for three different cleaved (001) surfaces of single crystal Sn$_{0.6}$In$_{0.4}$Te. 
		}
	}
	\label{fig:corelevels}	
\end{figure}

In Figure {\ref{fig:corelevels} we perform a UPS quantification analysis on bulk crystals of Sn$_{0.6}$In$_{0.4}$Te, grown from the same source material as the thin films in the manuscript, using a modified bridgemann technique \cite{Balakrishnan2013}. In contrast to the thin films, here the composition is uniform and hence the true stoichiometry ($x=0.40$) is easily established by energy dispersive x-ray spectroscopy (EDX) to within an uncertainty of 2\%. The same UPS quantification technique as employed in the manuscript \cite{CrossSectionComment} yields an average indium fraction of $x=0.56$, which is very close to the value initially obtained for Film 1 in the manuscript ($x=0.58$). The discrepancy between the UPS and EDX measurements suggests that the theoretically calculated cross sections should be slightly corrected. Rescaling the result to match the EDX measurement and then applying the same scaling to the film measurements in the manuscript yields new indium fractions of $x=0.41$ and $x=0.23$.